\documentclass[12pt]{article}
\usepackage{amsfonts}
\newcommand{\be}{\begin{equation}}
\newcommand{\ee}{\end{equation}}
\newcommand{\ba}{\begin{array}}
\newcommand{\ea}{\end{array}}
\newcommand{\bea}{\begin{eqnarray}}
\newcommand{\eea}{\end{eqnarray}}
\newcommand{\pro}{\partial}

\newcommand{\hn}{{\hat n}}

\newcommand{\dfrac}{\displaystyle\frac}
\newcommand{\nn}{\nonumber}

\begin{document}
\title
{Lorentz Gauge Gravity and Induced Effective Theories}

\author{S.-W. Kim $^1$ and D.G. Pak $ ^2$}
\date{}
\maketitle
\begin{center}{(1) School of Physics, College of Natural Sciences, 
Seoul National University, Seoul 151-742, Korea\\
(2) Center for Theoretical Physics, Seoul National
University, Seoul 151-742, Korea, and \\
Institute  of  Applied Physics,  Uzbekistan  National  University, 
Tashkent  700-095, Uzbekistan}
\end{center}

{\bf Abstract}: We develop the gauge approach 
based on the Lorentz group to the gravity with torsion.
With a  Lagrangian quadratic in curvature we show that
the Einstein-Hilbert action can be induced from a simple gauge 
model due to quantum corrections of torsion via formation of a gravito-magnetic
condensate. An effective theory of cosmic knots at Planckian scale is proposed.

\vskip 2mm
{\bf 1. Utiyama-Kibble-Sciama gauge approach to gravity}
\vskip 2mm

The gauge approach to gravity based on Lorentz and Poincare group
was proposed in \cite{uti} and later
was developed  
in many studies (see refs. in \cite{hehl}).
The Lorentz gauge models 
were further studied by Carmeli \cite{carmeli1}.
The possibility of inducing the Einstein gravity 
via quantum corrections was considered by many physicists
in various models
\cite{ogiev}.
In most of these models the Einstein-Hilbert term is induced by
quantum corrections due to interaction with matter fields.

In the present article we propose a simple gauge model of quantum
gravity based on the Lorentz group as a structural group. 
In the framework of this gauge model we demonstrate that 
even in a pure quantum gravity case with torsion the Einstein-Hilbert action can be
induced due to the quantum dynamics of torsion
via formation a non-trivial vacuum with a gravito-magnetic 
condensate.
We develop the gauge approach to
the gravity by suggesting
that {\it the torsion represents exactly 
the dynamical variable of quantum gravity.
Moreover, we conjecture that 
the torsion can be confined and exists intrinsically as a quantum object, 
and its quantum dynamics manifests itself by
inducing the Einstein-Hilbert theory as an effective theory of quantum
gravity}. 

We start with the formalism of the Lorentz 
gauge model along the lines proposed in
\cite{uti}.
The vielbein $e_a^\mu$ is treated as a 
fixed background field which obtains the dynamical content
after inducing the Hilbert-Einstein term in the effective theory. 
The covariant derivative with respect to the Lorentz
structural group is defined in a standard manner 
\bea
D_a=e_a^\mu (\pro_\mu + {\bf A}_\mu) ,
\eea   
where ${\bf A}_\mu\equiv A_{\mu cd} \Omega^{cd}$ is a general
affine connection taking values in Lorentz Lie algebra. 
The original Lorentz gauge transformation is the following
\bea
&& \delta e_a^\mu = \Lambda_a^b e_b^\mu, ~~~~~~~~~~
\delta {\bf A}_\mu=-\pro_\mu {\bf \Lambda} - [{\bf A}_\mu, {\bf \Lambda}],
\eea
where ${\bf \Lambda} \equiv \Lambda_{cd} \Omega^{cd}$.
One can split a general gauge connection $A_{\mu cd}$
into two parts, the background (classical) part and the quantum one. 
In what follows we will specify the classical background as one 
corresponding to
Riemanian space-time geometry
\bea
&&A_{\mu c}^{~~d} = \varphi_{\mu c}^{~~d} (e) + K_{\mu c}^{~~d} , \label{split}
\eea
where $K_{\mu c}^{~d}$
is a contorsion, and  $\varphi_{\mu c}^{~~d} $ is Levi-Civita spin connection
given in terms of the vielbein.
In the presence of contorsion we have two types of local 
symmetry transformations: \newline
(I) the classical, or background, gauge transformation
\bea
&& \delta {\bf \varphi}_\mu = -\pro_\mu {\bf \Lambda}-
                 [{\bf \varphi}_\mu,{\bf \Lambda}], ~~~~~~~~~
\delta {\bf K}_\mu = -[{\bf K}_\mu,{\bf \Lambda}],
\eea
(II) the quantum gauge transformation
\bea
&& \delta {\bf \varphi}_\mu=0,  ~~~~~~~~~~~~~~~~~~~
 \delta {\bf K}_\mu= - \hat D_\mu {\bf \Lambda}-[{\bf K}_\mu,{\bf \Lambda}],
\eea
where the background covariant derivative is defined with the help of
Levi-Civita connection $
\hat D_\mu =\pro_\mu +{\bf \varphi}_\mu$, ${\bf \varphi}_\mu
 \equiv \varphi_{\mu cd} \Omega^{cd}$.

Following the gauge principle as a guiding rule
we postulate the  gauge symmetry 
in the model under these two types of
transformations. The postulate restricts strongly
the admissible gauge invariants as possible candidates for the Lagrangian.
For instance, all terms quadratic in torsion (like
contact terms) are
forbidden since they 
spoil  the type II gauge invariance and by this the renormalizability
of the theory.

Let us remind the main lines
of Rieman-Cartan geometry (see, for ex., \cite{hehl}).
To define the derivative $D_\mu$ covariant
under the space-time diffeomorphisms 
one should
include a general Cristoffel symbol $\Gamma_{\mu \rho}^\nu$ 
\bea
&&  D_\mu V^\nu=\pro_\mu V^\nu +\Gamma_{\mu \rho}^\nu V^\rho.
\eea
The Cristoffel symbol is related to a general 
Lorentz connection $\gamma_{\mu a}^{~\,b}$ through the equation
$D_\mu e^{\rho a}= \pro_\mu e^{\rho a} +
\Gamma_{\mu \nu}^\rho e^{\nu a}-e^{\rho b} \gamma_{\mu b}^{~\,a}
=0$. This allows to convert space-time indices into Lorentz ones 
and vise versa by using the vielbein.
The contorsion is connected with torsion as follows 
\bea
&& K_{abc} = e_a^\mu K_{\mu bc} = -\dfrac{1}{2} (T_{abc}-T_{bca}+T_{cab}) .
\eea

Upon making the decomposition (\ref{split}) 
the curvature tensor is splitted into two parts
\bea
& R_{abcd}=\hat R_{abcd}+\tilde R_{abcd},  \nn \\
& \hat R_{abcd}=-\hat D_{[\underline a} \varphi_{\underline b] \underline c}^{
~\,\underline d}-\varphi_{[a|c}^{~~\,\,e}\varphi_{b]e}^{~\,d} , \nn \\
&\tilde R_{abcd}=-\hat D_{[\underline a} K_{\underline b] \underline c}^{
~\,\underline d}-K_{[a|c}^{~~e} K_{b]e}^{~\,d} ,
\eea
where the underlined indices stand for indices over which 
the covariantization is performed, and here we introduce a short notation 
for the antisymmetrization over indices $[a,b] = ab-ba$.

The classical action for a pure quantum
gravity in our approach contains the Maxwell type term quadratic in curvature
\bea
 S_{cl}&=& \dfrac{1}{4g^2} \int \sqrt {-g} d^4x  tr {\bf R}_{\mu \nu}^2=
-\dfrac{1}{4g^2} \int \sqrt {-g} d^4x R_{\mu \nu cd} R^{\mu \nu cd}, \label{L0}
\eea
where ${\bf R}_{\mu \nu} \equiv R_{\mu \nu cd} \Omega^{cd}$, and
we have written down explicitly
a new gravitational gauge coupling constant
$g$ corresponding to the Lorentz gauge group.
For brevity of notations we will use a
redefined contorsion which absorbs the coupling constant.
The same Lagrangian with the general Lorentz connection 
constructed from $SL(2,C)$ dyads and vielbeins was
considered by Carmeli \cite{carmeli1}.
It was demonstrated that the corresponding equations
of motion after projection with vielbein result in Newman-Penrose
form of Einstein-Hilbert equation in the vacuum.
Later Martellini and Sodano considered Carmeli's model
treating the connection as an independent quantity on vielbein
and proved the renormalizability of the model \cite{martell}. 

One should mention, since the Lorentz group is not compact
the classical Lagrangain leads to the Hamiltonian
which is not positively definite.
We adopt the point of view that even though 
the classical action (\ref{L0}) does not 
lead to a positively definite 
Hamiltonian, nevertheless, 
a consistent quantum theory can be formulated.
Since the canonical quantization method fails to handle our 
model we will apply the quantization scheme based on 
continual functional integration in Euclidean space-time.
Within this quantization scheme the quantum 
theory can be constructed since in the
Euclidean space-time
the Lorentz group is locally isomorphic to the product
of compact unitary groups 
$SU(2) \times SU'(2)$. 

\vskip 2mm
{\bf 2. Effective action}
\vskip 2mm

The general approach to derivation of the effective theory
is to integrate out all high energy (heavy mass)
modes while keeping light modes (massless or light particles).
Starting with the classical action (\ref{L0})
and imposing the gauge fixing condition $\hat D_\mu {\bf K}^\mu=0$ 
one can write down the effective action 
\bea
&&\exp \Big[iS_{eff} \Big] = 
\int
{\cal D} K_{\mu cd} {\cal D} {\bf c} {\cal D} \bar {\bf c} 
\exp \Big{\lbrace} \dfrac{}{} 
i\int \sqrt {-g} d^4x {\rm tr} \Big[ \dfrac{1}{4}  \hat {\bf R}^2_{\mu \nu}\nn \\
&+&\dfrac{1}{2} {\bf K}_{\mu} (g_{\mu \nu} \hat D \hat D - 
4 \hat {\bf R}_{\mu \nu}) {\bf K}_\nu
+\bar {\bf c} (\hat D \hat D) {\bf c}~\Big] \Big{\rbrace} ,
\eea
where ${\bf c}, \bar {\bf c}$ are Faddeev-Popov ghosts.
The formal expression for the one-loop effective action can be 
written in the form
\bea
 S_{eff} &=& S_{cl} -\dfrac{i}{2} Tr \ln [(g_{\mu \nu} (\hat D \hat D)_{ab}^{cd}-
     2 \hat R_{\mu \nu}^{~~ef} (f_{ef})_{ab}^{cd})] 
+ i Tr \ln [(\hat D \hat D)_{ab}^{cd}], \label{det1}
\eea
where $(f_{ef})_{ab}^{cd}$ are the structural constants of Lorentz Lie algebra.
The functional determinants in (\ref{det1}) are not well-defined 
in Minkowski space-time. As is known, the adding of
infinitesimal number factor $-i \epsilon$ to the 
bare Laplace operator in 
$\hat D \hat D$ is conditioned by the requirement 
of causality. The infinitesimal
addition $-i \epsilon$ defines uniquely the 
Wick rotation from Minkowski space-time to 
Euclidean one. In our case we should perform the 
Wick rotation in the base space-time and in the
tangent space-time both, so that the Lorentz group 
in Euclidean sector turns into the compact group 
$SO(4)\simeq SU(2)\times SU'(2)$. 
With this the functional integral becomes well-defined.
Certainly, there remains a problem of analytical continuation
of the final expressions from Euclidean space-time back to
Minkowski space-time. 

We have the following factorization for the Lie algebra valued 
curvature tensor 
\bea
&& R_{\mu \nu cd} \Omega^{cd} = -i(R_{\mu \nu}^i T^i+{R'}_{\mu \nu}^i T'^i),
\eea
where $T^i, T'^i$ are generators of the group $SU(2)\times SU(2)'$.

The functional determinants (\ref{det1})
are factorized into the direct product of $SU(2)$  
determinants, and the effective action
takes a simple form
\bea
S_{eff} &=& S_{cl} 
-\dfrac{i}{2} Tr \ln [(g_{\mu \nu} (\hat D \hat D)^{ij}-2\hat R_{\mu \nu}^k 
        \epsilon^{kij})] 
-\dfrac{i}{2} Tr \ln [(g_{\mu \nu} (\hat D' \hat D')^{ij}-2 \hat R'^k_{\mu \nu}
 \epsilon^{kij})] \nn \\
&& + i Tr \ln [(\hat D \hat D)^{ij}] 
+ i Tr \ln [(\hat D' \hat D')^{ij}] .
\eea
where all quantities corresponding to the group $SU(2)'$
are marked with apostrophe.

Notice that the curvature squared term contains a dual 
tensor ${\tilde {\hat R}} {}^{\mu\nu cd}$,
for instance,
\bea
(\hat R_{\mu \nu}^i)^2 &=&\dfrac{1}{8} (\hat R_{\mu \nu cd} \hat R^{\mu \nu cd} +
   \hat R_{\mu \nu cd} {\tilde {\hat R}} {}^{\mu\nu cd}) 
\equiv  \dfrac{1}{8} (\hat R^2 + \hat R \tilde {\hat R}), \nn \\
(\hat R'^i_{\mu \nu})^2 &=&\dfrac{1}{8} (\hat R_{\mu \nu cd} \hat R^{\mu \nu cd} -
   \hat R_{\mu \nu cd} {\tilde {\hat R}} {}^{\mu \nu cd}) 
\equiv \dfrac{1}{8} (\hat {R}^2 - \hat {R} \tilde {\hat {R}}).
\eea
For a constant background one can apply the
Schwinger's proper time 
method and $\zeta$-function regularization in full analogy with
the case of $SU(2)$ chromodynamics (QCD) \cite{cho3}. 
We will consider a constant homogeneous gravito-magnetic background 
field $H=\sqrt{\hat R_{\mu \nu cd}^2/2}$
which assumes that $\hat R_{\mu \nu cd} \tilde {\hat R}\,\!{}^{\mu \nu cd}=0$
in an appropriate coordinate frame.
For such gravito-magnetic background 
the final expression for the one-loop effective Lagrangian
is given by
\bea
&& {\cal L}_{eff} =-\dfrac{1}{2} H^2-\dfrac{11g^2}{48\pi^2}
       H^2(\ln\dfrac{gH}{\mu^2}-c), \\
&&c = 1-\dfrac{1}{2} -\dfrac{24}{11} \zeta(-1,\dfrac{3}{2})=1.29214...\,. \nn
\eea
With a proper
renormalization condition   
$\dfrac{\pro^2 V}{\pro H^2} \Big |_{H=\bar \mu^2}=\dfrac{1}{\bar g^2}$
one can obtain the renormalized effective potential 
\bea
&& V = \dfrac{1}{2 {\bar g}^2} H^2 + \dfrac{11}{48\pi^2}
       H^2(\ln\dfrac{H}{{\bar \mu}^2}-\dfrac{3}{2}).
\eea
One can check that the effective potential satisfies a
renormalization group  equation 
with the same $\beta$-function as in 
a pure $SU(2)$ Yang-Mills theory.

The minimum of the effective potential leads to a
gravito-magnetic condensate 
\bea
&& <H> = \bar \mu^2 \exp [-\dfrac{24 \pi^2}{11 \bar g^2}+1].
\eea

The presence of the minimum of the effective potential
does not guarantee that the corresponding new vacuum
is stable. The stability of the vacuum condensate 
even in the pure $SU(2)$ model of QCD
presents a long-standing problem, and its solution 
have passed through 
several controversal results since of the first
paper on that by Nielsen and Olesen \cite{niel}. 
Without clear evidence or at least a strong indication
to the vacuum stability one can not make any serious statement based
on existence of a non-trivial vacuum condensate.
Recently the progress in resolving that problem
in favor of stability of the magnetic vacuum
has been achieved in \cite{cho3}.
Moreover, it has been found recently
\cite{ff} that a stable classical configuration
made of monopole-antimonopole strings does exist in $SU(2)$
model of QCD providing a strong argument that a stable
magnetic vacuum can exist in QCD and, therefore, in 
our model of quantum gravity as well.

\vskip 2mm 
{\bf 3. Effective induced theories}
\vskip 2mm

Due to two types of gauge symmetries the condensate of torsion 
must vanish $<T_{abc}>=0$. It is possible that there is 
a deep analogy with QCD, and the torsion plays a
role of the off-diagonal (valence) gluon in QCD,
so that one can expect that the torsion can be confined. 
The presence of a stable gravito-magnetic condensate
generates a new scale in the theory, and 
one can also expect a non-vanishing vacuum averaged value for 
the curvature containing the torsion part 
\bea
<\tilde R_{abcd}> = M^2 (\eta_{ac} \eta_{bd}-\eta_{ad}\eta_{bc}).
 \label{assumpn}
\eea
The sign of the number factor $M^2$ is chosen positive since it
corresponds to the positive curvature space-time which can only be 
created due to quantum fluctuations.

Expanding the original Lagrangian near the
vacuum one obtains the Einstein-Hilbert Lagrangian
and the cosmological constant term in lower order approximation
(in units $\hbar=c=1$)
\bea
&&{\cal L}  = -\dfrac{1}{4g^2} (\hat R_{abcd}+\tilde R_{abcd})^2 
 = - \dfrac{1}{4 g^2} \hat R_{abcd}^2 - \dfrac{1}{16 \pi G} (\hat R
+2 \lambda) + ....
\eea
where the Newton constant $G$ and the cosmological constant $\lambda$
are defined by only one parameter, the renormalized coupling constant $\bar g$ 
at some scale $\bar \mu^2$ which supposed to be of order of Planckian
scale $10^{19} Gev$.

Certainly, the assumption (\ref{assumpn})
leads straightforward to a desired
induced Einstein-Hilbert term what was known very well before.
The most important point is how to make foundation
to that hypothesis. In our approach we put this assumption
on the real ground by the explicit calculation of the effective potential
and derivation of a stable classical 
vacuum configuration in $SU(2)$ Yang-Mills theory \cite{ff}.

Till now the numeric value of the 
gauge coupling $\bar g$ was not fixed, and it is a free
parameter in the theory. It is possible that
there are two phases corresponding to the strong 
and weak coupling constant. The existence of two phases in gravity was suggested
in \cite{polonyi} in a different approach.
In the strong coupling phase 
we can adjust the coupling constant $\bar g$ 
to $\bar g^2 \simeq 19$ to obtain the value
for $G$ close to the experimental value of the Newton constant. 
This provides also a large value for the cosmological constant
which is consistent with cosmological models containing the initial inflation 
at very early universe. 

It is interesting to 
consider the possibility of existence of a weak coupling 
phase with $\bar g^2/4 \pi < 1$. Using the experimental data for the 
vacuum energy density
$ \rho_v=\dfrac{2\lambda}{16 \pi G} =2\cdot 10^{-47} (Gev)^4$
one can find an appropriate value for the structure constant
$ \alpha_{\bar g} =\bar g^2/4 \pi=0.0123$. 
This value can be compared with the value  $\alpha_{SSGUT}\simeq 1/24$  
of the structure constant
in supersymmetric $SO(10)$ GUT model at unification scale 
$2 \times 10^{16} Gev$.
The same order of the structure constants $\alpha_g$ and $\alpha_{SSGUT}$
might be a hint to the origin of the supersymmetry
and its relation to quantum gravity.

Since the Lorentz group $SO(1,3)$ contains a maximal compact 
subgroup $SO(3)\simeq SU(2)$ we have the same homotopy structure
as in $SU(2)$ QCD, in particular,
the Hopf mapping
$\pi_3(SO(1,3)/SO(2)) ={\mathbb Z}$.
This suggests the existence of topological solitons
with non-trivial Hopf numbers like the knots
in Faddeev-Niemi-Skyrme model \cite{fadd}.
It has been shown that the generalized Faddeev-Niemi-Skyrme model
appears as an effective theory of QCD \cite{cho3}.
The derivation was based on 
Abelian decomposition of the $SU(2)$ gauge potential \cite{cho1}.
The essential part of this decomposition is represented by
a topological triplet $\hat n^i$ which parameterizes 
the coset $S^2 \simeq SU(2)/U(1)$. We can apply the results
obtained in $SU(2)$ QCD to our model in a case
of gravito-magnetic background $H$.
With this the effective Lagrangian corresponding to the
generalized Faddeev-Niemi-Skyrme model in gravity is given by \cite{cho3}
\bea
&{\cal L}_{eff} = -\dfrac{\mu^2}{2} (\pro_\mu \hn)^2 -\dfrac{1}{4} 
(\pro_\mu \hn \times \pro_\nu \hn)^2-\dfrac{\alpha_1}{4} 
(\pro_\mu \hn \cdot \pro_\nu \hn)^2
-\dfrac{\alpha_2}{2} (\hn\times\pro^2 \hn)^2,
\eea
where $\mu,\alpha_1,\alpha_2$ are parameters proportional to
vacuum averaging values of operator products of the magnetic potential
$\tilde C_\mu (\hn)$.
When the parameters $\alpha_{1,2}$ vanish the Lagrangian coincides 
with one of Faddeev-Niemi-Skyrme model \cite{fadd}, so that we expect 
the existence of cosmic knot solutions at Planckian scale with the mass
of order $M_{Planck}$. Recently the knot-like cosmic strings 
were considered in \cite{duan3}.

The detailed consideration of our results will be presented elsewhere.
\vskip 1mm
{\bf Acknowledgements}
\vskip 1mm
Authors thank Y.M. Cho for useful discussions.
One of authors (DGP) acknowledges E.I. Ivanov and B.M. Zupnik for
the kind hospitality during the workshop "SQS'05".

\end{document}